\documentclass[twocolumn,printnumbers,amsmath,amssymb,showpacs]{revtex4}
\usepackage{graphicx}% Include figure files
\usepackage{color}

\begin{document}
\title{Critical scalings and jamming in thermal colloidal systems}

\author{Lijin Wang}
\author{Ning Xu$^*$}

\affiliation{CAS Key Laboratory of Soft Matter Chemistry, Hefei National Laboratory for Physical Sciences at the Microscale $\&$ Department of Physics, University of Science and Technology of China, Hefei 230026, People's Republic of China.}

\date{\today}

\begin{abstract}
During the jamming of thermal colloids, the first peak of the pair distribution function shows a maximum height $g_1^{\rm max}$.  We find that $g_1^{\rm max}$ is accompanied by significant change of material properties and thus signifies the transition from unjammed to jammed glasses.  The scaling laws at $g_1^{\rm max}$ lead to scaling collapse of structural and thermodynamic quantities, indicating the criticality of the $T=0$ jamming transition.  The physical significance of $g_1^{\rm max}$ is highlighted by its coincidence with the equality of the kinetic and potential energy and the maximum fluctuation of the coordination number.  In jammed glasses, we find the strong coupling between the isostaticity and flattening of the density of vibrational states at the isostatic temperature scaled well with the compression.
\end{abstract}


\maketitle

As described by the seminal jamming phase diagram \cite{liu}, a packing of frictionless spheres interacting via repulsion undergoes the jamming transition at a critical point denoted as `J', which generalizes the jamming of colloids and granular materials  \cite{zhang,song,torquato,bi}.  At $T=0$, marginally jammed solids near Point J show particular critical scalings and length scales \cite{ohern,silbert1,wyart1,ellenbroek,olsson,xu1}.  Although recent theoretical work suggests that the jamming transition does not directly link to the glass transition in the hard sphere limit \cite{zhang,krzakala,berthier2,parisi}, the $T=0$ jamming transition still sheds some light on understanding the nature of glasses and amorphous solids \cite{xu1,xu2}.

Like the long-standing glass transition problem \cite{debenebetti,berthier1}, the structural similarities and absence of long range order during the jamming transition obscure proper order parameters to show diverging static correlation length, so the nature of the jamming transition still remains elusive.  Only recently, a structural vestige of the $T=0$ jamming transition has been observed in experiments and simulations of thermal colloidal systems \cite{zhang,cheng}.  At Point J, particles are just in contact, which induces a $\delta-$function of the first peak of the pair distribution function $g(r)$ \cite{silbert2,donev}.   At $T>0$, $g_1$, the height of the first peak of $g(r)$, reaches a maximum value $g_1^{\rm max}$ at a temperature dependent volume fraction (pressure).  In the $T=0$ limit, $g_1^{\rm max}$ diverges at Point J \cite{zhang}.

It remains an open question whether the emergence of $g_1^{\rm max}$ is merely the thermal vestige of the $T=0$ jamming transition.  If not, does it imply any physical significance in the formation of amorphous solids?  What features of the $T=0$ jamming transition would persist at $T>0$ and in what manner?  In this letter, we provide strong evidence via molecular dynamics simulations at low temperatures to confirm that the emergence of $g_1^{\rm max}$ contains rich and important physics.  It marks the transition from unjammed to jammed glasses by showing critical scalings,  recovering of some $T=0$ jamming features, and remarkable signs in the fluctuations and energy competition.  In jammed glasses, we find a novel isostatic temperature $T_i$ at which the isostaticity and flattening of the density of vibrational states coincide.  The isostatic temperature is scaled well with the compression.  Interestingly, the onset frequency of the plateau in the density of vibrational states at $T_i$ shows similar pressure dependence to marginally jammed solids at $T=0$.

Our systems are three-dimensional cubic boxes consisting of $N=1000$ spheres with the same mass $m$ \cite{zhang}.  Half of the spheres have a diameter of $\sigma$, while the other half have a diameter of $1.4\sigma$, to avoid crystallization.  Periodic boundary conditions are applied in all directions.  Particles $i$ and $j$ interact via repulsive potential $V_{ij}=\frac{\epsilon}{\alpha}\left( 1 - r_{ij}/\sigma_{ij}\right)^{\alpha}$ when their separation $r_{ij}$ is less than the sum of their radii $\sigma_{ij}$, and zero otherwise.
In this letter, we only show the results for harmonic repulsion ($\alpha=2$). The results for Hertzian repulsion ($\alpha=5/2$) are shown in the Supplementary Information.  We set the units of mass, energy, and length to be $m$, $\epsilon$, and $\sigma$, and the Boltzmann constant $k_B=1$.  The frequency is in the units of $\sqrt{\epsilon/m\sigma^2}$.  We perform molecular dynamics simulations at constant temperature and pressure using Gear predictor-corrector algorithm.  The density of vibrational states $D(\omega)$ is obtained from the Fourier transform of the velocity correlation function $C(t)=\left< \vec{v}(t)\cdot \vec{v}(0)\right>$ \cite{keyes}:
\begin{equation}
D(\omega)=\frac{1}{3T}\int_0^{\infty} C(t){\rm cos}(\omega t) {\rm d}t, \label{dos}
\end{equation}
where $\left< .\right>$ denotes the time and particle average.

%%%%%%%%%%%%%%%%%%%%%%%%%%%%%%%%%%%%%%%%%%%%%%%%%%%
\begin{figure}
\includegraphics[width=0.47\textwidth]{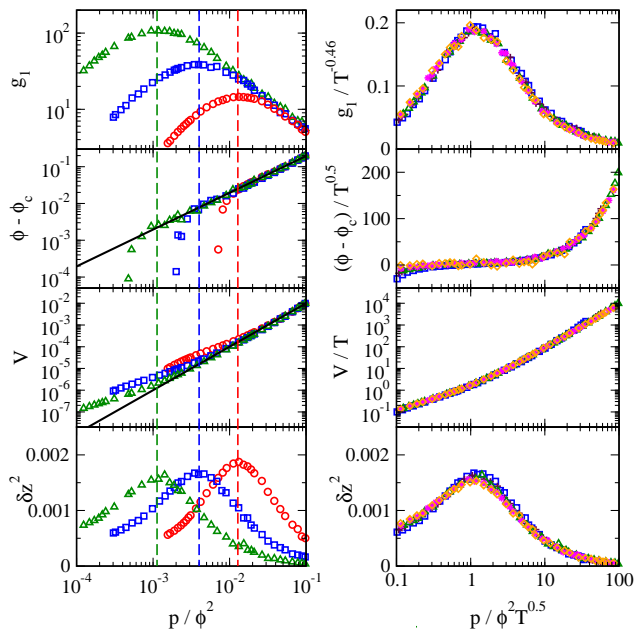}
\vspace{-0in}
\caption{\label{fig:fig1} (color online) Left panels: Pressure dependence of the height of the first peak of the pair distribution function $g_1$, volume fraction difference from the $T=0$ jamming transition $\phi-\phi_c$, potential energy per particle $V$, and mean square fluctuation of the coordination number $\delta z^2$ for systems with harmonic ($\alpha=2$) repulsion.  The black solid lines show the scalings in marginally jammed solids at $T=0$: $\phi-\phi_c\sim p/\phi^2$ and $V\sim (p/\phi^2)^2$.  The dashed lines mark the locations of $g_1^{\rm max}$.  Right panels: Scaling collapse of the quantities in the left panels.  The data are measured at $T=10^{-4}$ (red circles), $10^{-5}$ (blue squares), $4\times 10^{-6}$ (maroon pluses), $10^{-6}$ (green triangles), $4\times 10^{-7}$ (magenta stars), and $10^{-7}$ (orange diamonds).
}
\end{figure}
%%%%%%%%%%%%%%%%%%%%%%%%%%%%%%%%%%%%%%%%%%%%%%%%%%%

Figure~\ref{fig:fig1} provides sufficient information to point out the physical significance of the structural signature $g_1^{\rm max}$.  The left panels of Fig.~\ref{fig:fig1} indicate that along with the emergence of $g_1^{\rm max}$ at a crossover pressure $p_j$ multiple quantities undergo qualitative changes.  In a wide range of temperatures, both the volume fraction difference from Point J $\phi -\phi_c$ and potential energy per particle $V$ show distinct pressure dependence on two sides of $p_j$.  At $p>p_j$, some of the critical scalings in marginally jammed solids at $T=0$ are recovered:
\begin{equation}
\phi-\phi_c\sim \left(p/\phi^2\right)^{1/(\alpha-1)}, \label{phi_p}
\end{equation}
\begin{equation}
V \sim (\phi-\phi_c)^{\alpha}\sim \left(p/\phi^2\right)^{\alpha / (\alpha - 1)}, \label{V_p}
\end{equation}
where the critical volume fraction at Point J $\phi_c=0.649\pm 0.002$ for both harmonic ($\alpha=2$) and Hertzian ($\alpha=2.5$) repulsions.  We correct the widely accepted scaling $p\sim \left( \phi-\phi_c \right)^{\alpha-1}$ \cite{ohern} to Eq.~(\ref{phi_p}) which can be simply derived from Eq.~(\ref{V_p}) and $p=\phi^2\frac{dV}{d\phi}$, because away from $\phi_c$ the variation of $\phi$ must be considered to interpret the data correctly.  At $p<p_j$, however, the jamming-like scalings break down, indicating the notable change of the material properties across $p_j$.

In addition to Eqs.~(\ref{phi_p}) and (\ref{V_p}), the average coordination number per particle $z$ is scaled well with the compression at $T=0$: $z-z_c\sim (\phi-\phi_c)^{1/2}\sim (p/\phi^2)^{1/2(\alpha-1)}$, where $z_c=2d$ is the isostatic value with $d$ the dimension of space.  Isostaticity ($z=z_c$) is one of the most special features of the $T=0$ jamming transition at Point J.  It controls the unusual vibrational properties of marginally jammed solids and realistic glasses \cite{ohern,silbert1,wyart1,xu2}.  Its role in vibrations of jammed glasses at $T>0$ will be discussed later.   At $T>0$, the thermal motion breaks particle contacts frequently, so the $T=0$ scaling of $z$ no longer exists in the vicinity of $g_1^{\rm max}$.  We observe that $z_j$, the coordination number at $p_j$ is less than $z_c$ and gradually decreases with decreasing the temperature.  Meanwhile, the slope $(\frac{dz}{dp})_{z_j}$ grows rapidly to infinity in the $T=0$ limit.  This is consistent with the discontinuous jump of the coordination number from $0$ to $z_c$ at Point J \cite{ohern}.

Although the coordination number does not maintain the $T=0$ scaling at $p>p_j$, its fluctuation $\delta z^2$ is interestingly correlated to $g_1^{\rm max}$ by peaking at $p_j$, as shown in the left bottom panel of Fig.~\ref{fig:fig1}.  It is natural to imagine that the fluctuation of particle contact is smaller in more solid-like systems.  The peak in $\delta z^2$ indicates that systems at $p>p_j$ is more rigid than those at $p<p_j$.  Furthermore, the peak value of $\delta z^2$ is almost independent on the temperature at low temperatures, implying the singularity of Point J, since $\delta z^2=0$ at $T=0$.

A close look at the potential energy unveils another interesting phenomenon: $V\approx \frac{3}{2}T$ at $p_j$.  At $p>p_j$, the potential energy dominates the kinetic energy, so that the thermal motion is not strong enough to cause significant configuration change of jammed states and the $T=0$ scalings thus hold.  At $p<p_j$, the kinetic energy wins.  The thermal motion has noticeable effects on the system performance.  Therefore, the emergence of $g_1^{\rm max}$ and the corresponding change of material properties do not happen by chance, but have explicit physical origins.

At $T=0$, the height of the first peak of the pair distribution function $g_1$ is inversely scaled with $\phi - \phi_c$ when the system is compressed away from Point J, because $g_1$ is inversely proportional to the particle overlap which grows linearly with $\phi-\phi_c$ \cite{zhang,silbert2}.  It has been shown that $T_j$, the temperature at $p_j$ is scaled well with $\phi-\phi_c$ \cite{zhang}:
\begin{equation}
T_j\sim (\phi-\phi_c)^{\alpha}\sim \left( p/\phi^2\right)^{\alpha / (\alpha - 1)}. \label{Tj}
\end{equation}
Neglecting the thermal effects on $g_1$ from $T=0$ to $T_j$, we can easily obtain $g_1^{\rm max} \sim (\phi-\phi_c)^{-1}\sim T^{-1/\alpha}$.  However, thermal and nonlinear effects still cause visible deviation from the prediction.  As shown in the Supplementary Information,
\begin{equation}
g_1^{\rm max} \sim (\phi -\phi_c)^{-\gamma} \sim \left( p/\phi^2\right)^{-\gamma/(\alpha-1)}, \label{g1_p}
\end{equation}
where $\gamma=0.90\pm0.02$.

Equations~(\ref{phi_p})$-$(\ref{g1_p}) imply the criticality of Point J viewed in thermal colloidal systems.  By scaling all the structural and thermodynamic quantities in the left panels of Fig.~\ref{fig:fig1} using Eqs.~(\ref{phi_p})$-$(\ref{g1_p}), we obtain excellent scaling collapse of the original low-temperature data.  As shown in the right panels of Fig.~\ref{fig:fig1}, there are continuous scaling functions covering data on both sides of $p_j$:
\begin{equation}
\xi = T^{\nu_{\xi}} f_{\xi} \left( \frac{p}{\phi^2T^{(\alpha - 1)/\alpha}}\right), \label{scaling}
\end{equation}
where $\xi$ denotes $g_1$, $\phi-\phi_c$, $V$, and $\delta z^2$ with $\nu_{\xi}=-\gamma/\alpha$, $1/\alpha$, $1$, and $0$ respectively.  During the writing of this letter, we notice that very recent approaches based on mean field theory has predicted similar scaling collapse \cite{berthier3,otsuki}.  Here we explicitly obtain Eq.~(\ref{scaling}) directly from the scalings present at $p_j$ without making any assumptions.  Furthermore, the scaling collapse of $\delta z^2$ is not able to predict from the mean field theory.   Our study here gives explicit physical origin and meaning of the scaling collapse, and highlights further the physical significance of $g_1^{\rm max}$.  Since the scaling collapse is simply derived from the scaling laws at $p_j$ which approaches Point J in the $T=0$ and $p=0$ limit, our finding here strongly supports the criticality of Point J.  The scaling collapse conveys important information of the low-temperature properties of the model systems in the vicinity of Point J, not just trivially reflecting the behaviors at $T=0$.

From the above discussions about Fig.~\ref{fig:fig1}, we can conclude that the emergence of $g_1^{\rm max}$ is not trivially the thermal vestige of the $T=0$ jamming transition.  It is plausible to state that $g_1^{\rm max}$ is the structural signature of the jamming transition from unjammed to jammed glasses at $T>0$.

%%%%%%%%%%%%%%%%%%%%%%%%%%%%%%%%%%%%%%%%%%%%%%%%%%%
\begin{figure}
\includegraphics[width=0.47\textwidth]{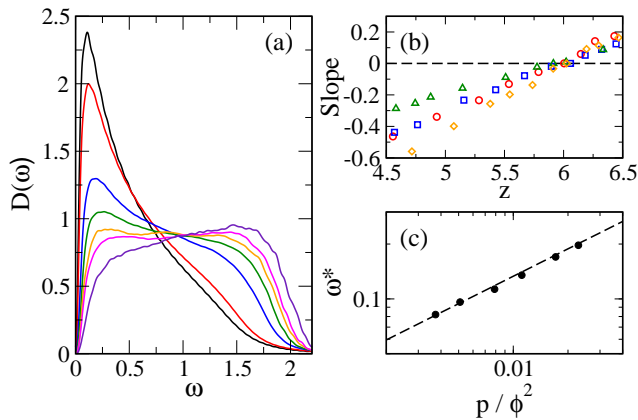}
\vspace{-2.in}
\caption{\label{fig:fig2} (color online) (a) Density of vibrational states $D(\omega)$ for systems with harmonic ($\alpha=2$) repulsion at $T=10^{-5}$.  The pressures are $0.0008$, $0.001$, $0.002$, $0.003$, $0.004$, $0.005$, and $0.009$ from the black to violet curves.  The crossover pressure $p_j$ is around $0.0015$.  (b) Slope of $D(\omega)$ in the intermediate frequency regime versus the coordination number $z$ for $T=10^{-4}$ (red circles), $10^{-5}$ (blue squares), $10^{-6}$ (green triangles), and $10^{-7}$ (orange diamonds).  The dashed line shows where the flattening of $D(\omega)$ happens. (c) Frequency $\omega^*$ close to the onset of the plateau in $D(\omega)$ with $D(\omega^*)\approx 0.66$ measured at $z=z_c$.  The dashed line shows the $T=0$ scaling, $\omega^*\sim (p/\phi^2)^{1/2}$.
}
\end{figure}
%%%%%%%%%%%%%%%%%%%%%%%%%%%%%%%%%%%%%%%%%%%%%%%%%%%

As mentioned above, the $T=0$ scaling of the coordination number $z$ is violated in jammed glasses near $p_j$.  At $p_j$, $z<z_c$ although $\phi>\phi_c$.  At fixed temperature, $z$ increases with the pressure, so isostaticity ($z=z_c$) would occur at $p>p_j$.  At $T=0$, isostaticity determines the special vibrational properties of marginally jammed solids \cite{silbert1,wyart1,xu1}.  For instance, the density of vibrational states $D(\omega)$ possesses a low-frequency plateau at Point J, which shrinks with increasing $\phi-\phi_c$ and eventually disappears \cite{silbert1}.  The characteristic frequency of the onset of the plateau is linearly scaled with $z-z_c$ for jammed systems with harmonic repulsion \cite{silbert1,wyart1}.  It is thus interesting to know if there are any connections between the isostaticity and vibration in jammed glasses at $T>0$.

Figure~\ref{fig:fig2}(a) shows the evolution of $D(\omega)$ measured from Eq.~(\ref{dos}) with the change of the pressure at fixed temperature.  Notice that all the systems shown here are glassy with the relaxation time much longer than the simulation time window.  At low pressures, there is a low-frequency peak which grows up with decreasing the pressure.  The same trend has also been observed in hard sphere systems \cite{brito}.  This low-frequency peak indicates the aggregation of soft modes, so the system with a higher peak would be less stable.  At $p_j$, this soft mode peak is still pronounced, in consistent with the fact that $z<z_c$.  At $p>p_j$, the peak is progressively suppressed with the pressure.  Interestingly, a low-frequency plateau is eventually formed in $D(\omega)$, which is also shown in a recent experiment of thermal colloids \cite{tan}.  Afterwards, the pressure dependence of $D(\omega)$ looks similar to that at $T=0$.

We estimate the average slope of $D(\omega)$ in the intermediate frequency regime at different temperatures, as shown in Fig.~\ref{fig:fig2}(b).  It is interesting that the flattening of $D(\omega)$ happens approximately when $z=z_c$ at the isostatic temperature $T_i$.  This striking finding indicates that the flattening of $D(\omega)$ is strongly coupled to the isostaticity, not only at Point J, but also in thermal colloidal systems with repulsion.  At $T_i$, although $z=z_c$ and there is a plateau in $D(\omega)$, the onset frequency $\omega^*$ of the plateau still relies on the pressure.  Figure~\ref{fig:fig2}(c) shows that the $T=0$ scaling of jammed solids with harmonic repulsion, $\omega^*\sim(\phi-\phi_c)^{1/2}\sim (p/\phi^2)^{1/2}$ \cite{silbert1,wyart1}, is still valid at $T_i$.

Figure~\ref{fig:fig2} unveils the complicated and puzzling effects of the thermal motion on the vibration.  Notice that the reference state at $T_i$, i.e. the configuration with time-averaged particle positions, still has a coordination number larger than $z_c$, although in time average $z=z_c$.  Isostaticity induced by thermal fluctuations only controls the flattening of $D(\omega)$.  The compression, or equivalently the reference state, may still determine some vibrational properties of jammed glasses.  Here we have the first attack on this interesting issue.  Further studies are needed to better understand the thermal effects.

\begin{figure}
\includegraphics[width=0.45\textwidth]{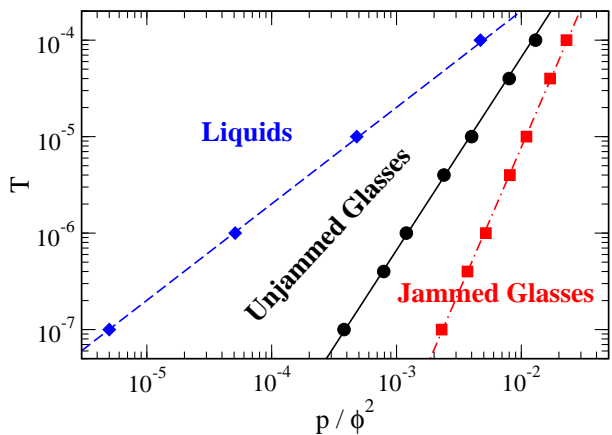}
\vspace{-2.4in}
\caption{\label{fig:fig3}  (color online) Phase diagram of systems with harmonic ($\alpha=2$) repulsion.  The glass transition temperature (blue diamonds), jamming transition temperature (black circles), and isostatic temperature (red squares) are plotted together as a function of $p/\phi^2$.  The lines are power-law fits to the data: $T_g\sim p/\phi^2$ (blue dashed), $T_j\sim (p/\phi^2)^2$ (black solid), and $T_i\sim (p/\phi^2)^3$ (red dot-dashed).
}
\end{figure}

The isostatic temperature $T_i$ is scaled well with the compression:
\begin{equation}
T_i\sim \left(\phi-\phi_c \right)^{\alpha + 1}\sim \left( p/\phi^2\right)^{(\alpha+1)/(\alpha-1)}.
\end{equation}
This scaling may be understandable from the competition between kinetic energy and potential energy to break extra contacts of the reference states beyond isostaticity.  Together with Eq.~(\ref{Tj}) and a recent observation of the linear pressure dependence of the glass transition temperature at which the relaxation time or viscosity diverges, $T_g\sim p$ \cite{xu3}, we are able to plot the phase diagram in Fig.~\ref{fig:fig3} to clarify the connections between different transitions.  At low temperatures, the glass transition, jamming transition, and isostaticity happen in sequence when the pressure (volume fraction) increases.  We have already known that there is a gap between the glass transition and jamming transition \cite{zhang,jacquina}.  Here we introduce the new isostaticity line which converges to Point J with the jamming transition in the $T=0$ limit, but likely on the two limits of Point J since $z$ is different on these two lines and discontinuous at Point J.  Figure~\ref{fig:fig3} demonstrates the complexity in the formation of amorphous solids.  Even in glasses, there are complicated thermodynamic and vibrational behaviors beyond our expectation, which requires intensive studies to sort through.

%%%%%%%%%%%%%%%%%%%%%%%%%%%%%%%%%%%%%%%%%%%%%%%%%%%
%%%%%%%%%%%%%%%%%%%%%%%%%%%%%%%%%%%%%%%%%%%%%%%%%%%

The findings discussed in this letter are limited to low temperatures and pressures where the influence of Point J is still sensible.  As shown in Fig.~\ref{fig:fig3}, if all the critical scalings hold at high temperatures, the three lines eventually get across.  Recent studies have shown that the glass transition temperature drops with increasing the pressure at sufficiently high pressures \cite{berthier4}, so nothing at high temperatures and pressures is predictable from the low-temperature and low-pressure properties.  Beyond the regime concerned here, we will enter a new territory with completely new deep jamming pictures \cite{zhao}.

We thank Emily S. C. Ching, Andrea J. Liu, Sidney R. Nagel, Yair Shokef, Peng Tan, Matthieu Wyart, and Lei Xu for helpful discussions.  We especially thank Ludovic Berthier for drawing us attention to related work and critical reading of the manuscript.  This work is supported by National Natural Science Foundation of China No. 91027001 and 11074228, National Basic Research Program of China (973 Program) No. 2012CB821500, CAS 100-Talent Program No. 2030020004, and Fundamental Research Funds for the Central Universities No. 2340000034.

\bf{Supplementary Information}

%%%%%%%%%%%%%%%%%%%%%%%%%%%%%%%%%%%%%%%%%%%%%%%%%%%
\begin{figure}[!htbp]
\vspace{0.1in}
\includegraphics[width=0.35\textwidth]{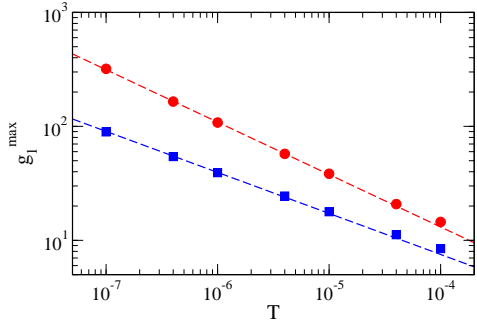}
\vspace{-0.2in}
\caption{\label{fig:figs1} Temperature dependence of the maximum value of the first peak of the pair distribution function, $g_1^{\rm max}$.  The red circles and blue squares are for systems with harmonic ($\alpha=2$) and Hertzian ($\alpha=2.5$) repulsions, respectively.  The red and blue dashed lines show the power law scalings: $g_1^{\rm max}\sim T^{-0.46}$ and $g_1^{\rm max}\sim T^{-0.36}$ ($g_1^{\rm max}\sim T^{-(0.90\pm0.02)/\alpha}$), respectively.}
\end{figure}
%%%%%%%%%%%%%%%%%%%%%%%%%%%%%%%%%%%%%%%%%%%%%%%%%%%

%%%%%%%%%%%%%%%%%%%%%%%%%%%%%%%%%%%%%%%%%%%%%%%%%%%
\begin{figure}[!htbp]
\includegraphics[width=0.35\textwidth]{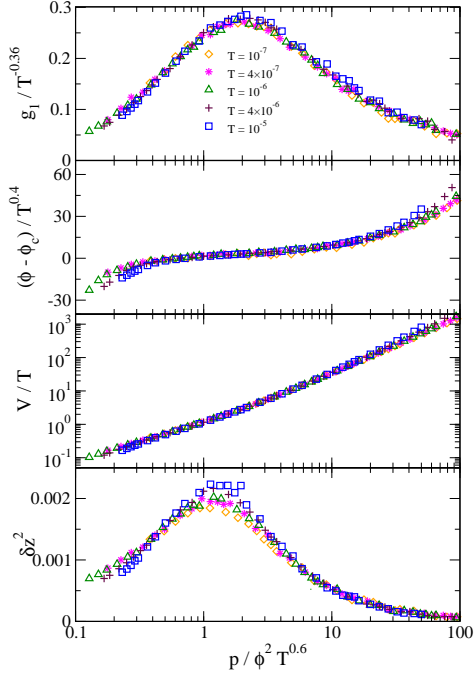}
\caption{\label{fig:figs2}  Scaling collapse of the height of the first peak of the pair distribution function $g_1$, volume fraction difference from the $T=0$ jamming transition $\phi-\phi_c$, potential energy per particle $V$, and mean square fluctuation of the coordination number $\delta z^2$ for systems with Hertzian ($\alpha=2.5$) repulsion.
}
\vspace{-2in}
\end{figure}
%%%%%%%%%%%%%%%%%%%%%%%%%%%%%%%%%%%%%%%%%%%%%%%%%%%

%%%%%%%%%%%%%%%%%%%%%%%%%%%%%%%%%%%%%%%%%%%%%%%%%%%
\begin{figure}[!ht]
\includegraphics[width=0.35\textwidth]{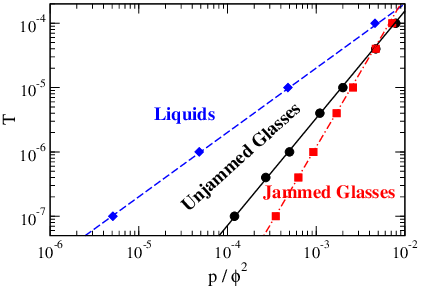}
\vspace{-0.1in}
\caption{\label{fig:figs3}  Phase diagram of systems with Hertzian ($\alpha=2.5$) repulsion.  The glass transition temperature (blue diamonds), jamming transition temperature (black circles), and isostatic temperature (red squares) are plotted together as a function of  $p/\phi^2$.  The lines are power-law fits to the data: $T_g\sim p/\phi^2$ (blue dashed), $T_j\sim (p/\phi^2)^{5/3}$ (black solid), and $T_i\sim (p/\phi^2)^{7/3}$ (red dot-dashed).
}
\end{figure}
%%%%%%%%%%%%%%%%%%%%%%%%%%%%%%%%%%%%%%%%%%%%%%%%%%%

\end{document}